\def\year{2020}\relax
\documentclass[letterpaper]{article}
\usepackage{multirow}
\usepackage{subfig}
\usepackage{aaai20}
\usepackage{times}
\usepackage{helvet}
\usepackage{courier}
\usepackage[hyphens]{url}
\usepackage{graphicx}
\usepackage{verbatim}

\title {On the Explanation of Similarity for Developing and Deploying CBR Systems}

\author{Kerstin Bach\\
Department of Computer Science, \\
Norwegian University of Science and Technology\\
\url{https://www.ntnu.edu/idi}\\
kerstin.bach@ntnu.no
\And
Paul Jarle Mork\\
Department of Public Health and Nursing, \\
Norwegian University of Science and Technology\\
\url{https://www.ntnu.edu/ism}\\
paul.mork@ntnu.no
}

\def\mcbr{\textsc{myCBR}}

\def\rapi{Rest API}

\begin{document}
\maketitle

\begin{abstract}
During the early stages of developing Case-Based Reasoning (CBR) systems the definition of similarity measures is challenging since this task requires transferring implicit knowledge of domain experts into knowledge representations. While an entire CBR system is very explanatory, the similarity measure determines the ranking but do not necessarily show which features contribute to high (or low) rankings. 
In this paper we present our work on opening the knowledge engineering process for similarity modelling. This work present is a result of an interdisciplinary research collaboration between AI and public health researchers developing e-Health applications. During this work explainability and transparency of the development process is crucial to allow in-depth quality assurance of the by the domain experts.

\end{abstract}

\section{Introduction}
Case-Based Reasoning (CBR) systems utilize previous experience in form of problem-solution pairs (cases) to solve new problems by matching the problem to its closest, most similar case \cite{AamodtPlaza1994}. During the retrieval phase of a CBR system the case representation and similarity measures are crucial to find solutions that are most relevant to a given problem, while in reuse the modifications of the solution are undertaken to better suit the problem description.

While CBR is often described as an open box and explainable AI (XAI) method \cite{Sormo2005Expl,Weber2018,Leake2005}, the similarity assessment usually provides a shallow explanation in form of a similarity score. The interpretation of the similarity score, however, can be challenging and especially during the development of CBR systems discussions between experts and knowledge engineers need to be facilitated. From our experience, the more explanatory the reasoning process is, the better the knowledge representation and refinements from expert become. Especially considering the Knowledge Containers \cite{Richter1995,Ganesan2018IJCAI}, knowledge engineers have four different types of knowledge that interplay with each other and hence influence the selection and ranking of cases. When the knowledge engineer or CBR expert does not have all available domain knowledge, collaboration with domain experts becomes increasingly important. Being able to explain the reasoning process within the domain with relevant cases helps to increase trust in the application by the expert as well as it leverages better evaluation in real-world setting and therewith detection of faults only.

In \cite{JaiswalBach2019,VermaBachMork2018} we have introduced methods that support the knowledge modeling process for CBR systems. In this paper we focus on methods, especially visualizations, that explain the reasoning process -- especially the similarity-based retrieval used in CBR systems. In the remainder of the paper we will present an overview of previous work of CBR and XAI. We will look into methods, applications and tools that contribute to enable explainable. In section \ref{sec:data} we will introduce a reference dataset for the remainder of the paper. We will present the current possibilities of explaining traditional machine learning methods before, in section \ref{sec:method} we show how similarity measures for the sample dataset can be defined. In section \ref{sec:vis} we will introduce visualizations that allow a better understanding of the reasoning process. The final sections summarizes our work and gives an outlook on the next steps.

\section{Related Work}
\label{sec:rw}

The explanatory capabilities of CBR has been addressed by researchers throughout the life cycle of the field. Especially the work by Leake \cite{Leake01} presents a general framework discussing issues that need to be addressed for explanations. Further on, \cite{Sormo2005Expl} present detailed explanation goals for a system: transparency, justification, relevance, conceptualization, and learning. Both works emphasize that not everything needs to be explained and that the context of an explanation needs to be taken into account.

Another aspect of explanations discussed in \cite{Massie2004} is that users on the one hand gain confidence in a system that provides correct results, but confidence is also improved when the decision making process is understood and deficiencies can be identified and resolved. This view on similarity measures and their role during the retrieval will be addressed by our work later in this paper.

In previous work, the explanation of the case base content has been discussed by Smyth and McKenna \cite{Smyth2000,McKenna2001} and they suggested visualisations of case base content and making changes to the case base explicit to the user. In more recent work, the authors of \cite{keane2019case} present how CBR has been used to explain neural networks, which describes another branch of XAI. 


While the theoretical concepts are important to move a field forward, their implementation in practice allows the research field to grow and attract others. CBR tools have been developed since the very beginning of the CBR research
activities. The most general CBR tools developed and provided as bundled or open
source software are COLIBRIStudio (and their predecessors COLIBRI, jCOLIBRI)
\cite{DIAZAGUDO200768}, CBRworks \cite{Schulz99cbr-works} and its successor
\mcbr{\cite{stahl2008rapid}}. Furthermore there are more specific CBR tools
targeting certain domains or case representations. For process-oriented CBR, the
Collaborative Agent-based Knowledge Engine (CAKE)
\cite{bergmann_collaborative_2014} has been introduced, while CREEK
\cite{Aamodt04} is a tool for knowledge-intense CBR and (B)EAR
\cite{Jalali2015,Jalali2016} focuses on the adaptation in CBR systems.

\mcbr{\footnote{http://mycbr-project.org}}, which is used and extended in this work, was developed by German Research Center for Artificial Intelligence (DFKI) and has
been introduced as rapid prototyping tool for research and industrial
applications. 
Recently the tool was generalized and provides a \rapi{} for more flexible interaction with the engine and it's components \cite{BachEtAl2019}. Each component in \mcbr~is explainable, which allows a deep integration of explanations in knowledge modelling, but also reasoning \cite{Bahls2007}. 

\section{Example Dataset}
\label{sec:data}

CBR researchers have in the past very closely been collaborating with the health care domain due to its explanatory and transparent nature \cite{GonzalezLopezBlobel2013}. Since our work is also linked to patient-centered e-Health applications, we will introduce an open dataset from this domain to be used as a running example in this paper.

In the following we will describe different aspects of similarity modelling using an open dataset from the UCI Machine Learning Library \cite{Dua:2019} containing 768 female patients of Pima Indian heritage. The dataset was introduced by \cite{PimaDiabetesDataset} and has been provided by the National Institute of Diabetes and Digestive and Kidney Diseases to diagnostically predict whether or not a patient has diabetes, based on certain diagnostic measurements included in the dataset. 

\begin{figure}[h]
  \centering
  \includegraphics[width=0.47\textwidth]{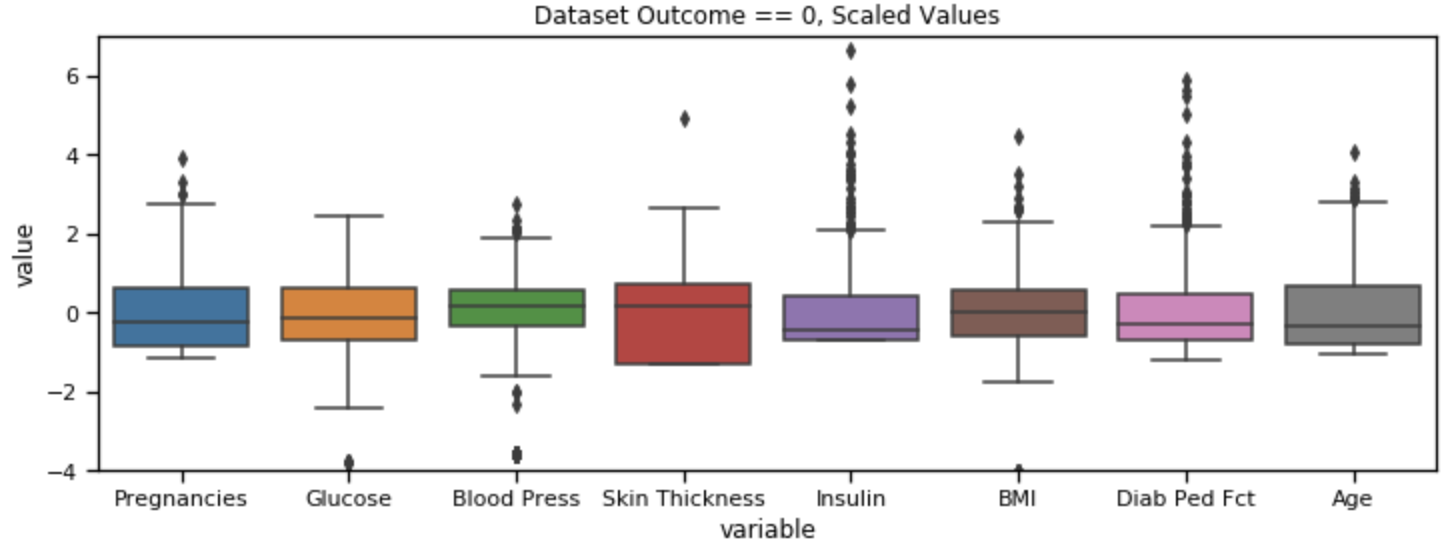}
  \caption{Value Distribution in the Pima Diabetes Dataset}
  \label{fig:data}
\end{figure}

While the dataset describes the characteristics of patients included in the cohort, the main usage of the dataset is to classify whether a patient has diabetes or not. Figure \ref{fig:data} shows the value distribution within the diabetes dataset. Even when grouping them by outcome, the individual distributions are very similar, which shows that there is no apparent feature indicating the outcome. 

\begin{figure}[h]
  \centering
  \includegraphics[width=0.47\textwidth]{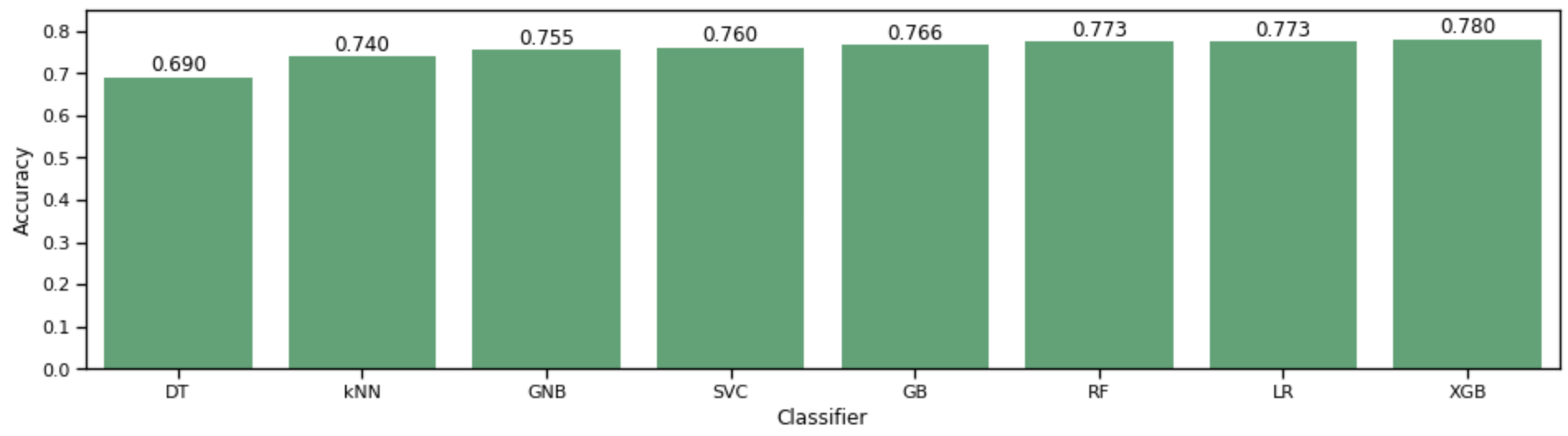}
  \caption{Benchmark of 8 classifiers on the diabetes dataset (kNN:  k-Nearest Neighbour, SVC: Support Vector Machine , LR: Logistic Regression, DT: Decision Tree, GNB: GaussianNB, RF: Random Forest, GB: Gradient Boosting, XGB: XGBoost)}
  \label{fig:ml-cv}
\end{figure}

As a reference point, we implemented standard machine learning approaches to carry out this task\footnote{The authors will make the code for training these classifiers available for the final version.} and as it can be seen in Figure \ref{fig:ml-cv} with only basic tuning of the parameters the average accuracy only reaches about $0.78\%$ for the best performing classifier on a 10-fold cross-validation. 

Many of these methods would also allow to provide information about the feature importance for the classification, but this information still lacks detail on how the model is build and why certain features are chosen.

\section{Similarity Modeling and Retrieval}
\label{sec:method}
The similarity modelling in a CBR system can vary from as simple as a kNN to knowledge-intense graph-based representations as presented in \cite{Aamodt04}. CBR tools, such as \mcbr{} that has been used in this work provide predefined similarity measure that cover mostly symbolic and numeric value ranges. In the example set of this paper we only have numerical values. The main similarity measure method of \mcbr{} is defining similarity according the local-global-principle\cite{Richter1995}. Thereby similarity measures are defined as an amalgamation function, such as a weighted sum, which defines as local similarity measures the relationship of attribute values and the global similarity the weighted sum of the local similarities.

\subsection{Defining Similarity Measures}
Modelling similarity measure can be done automatically using neural networks \cite{gabel2015ann}, feedback from users \cite{Stahl2005} or from the training data \cite{Mathisen2019}. 

\begin{figure}[h]
  \centering
  \includegraphics[width=0.47\textwidth]{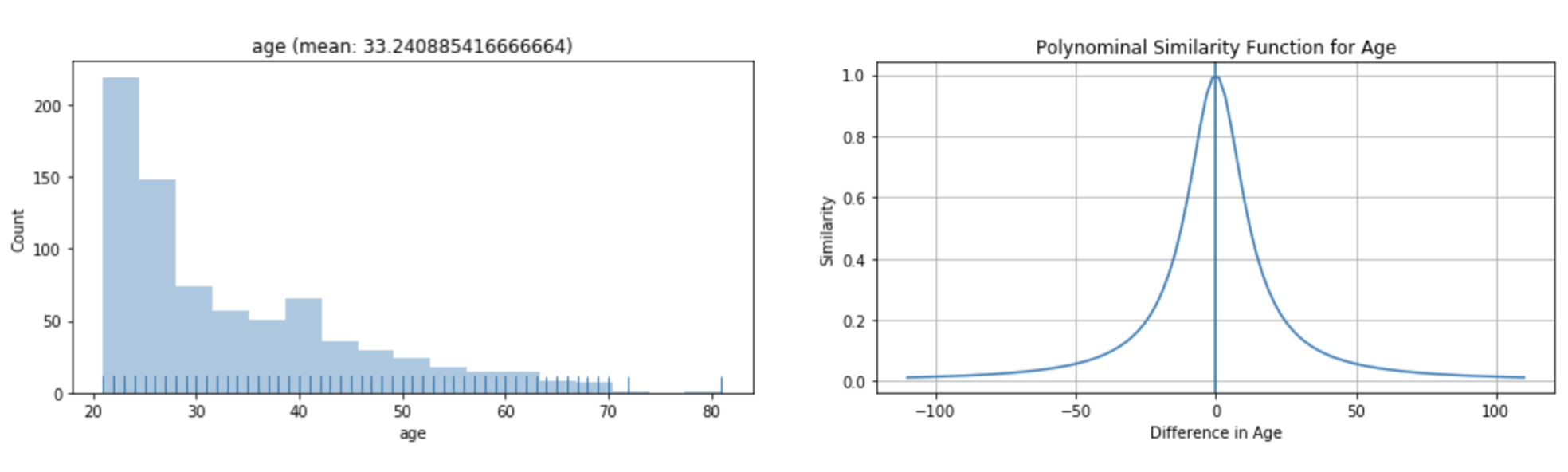}
  \caption{Example of a polynominal similarity measure for the attribute \textit{Age}: data distribution on the left and polynominal function on the right}
  \label{fig:age-sim}
\end{figure}

Figure \ref{fig:age-sim} shows a data-driven or manual (expert-based) definition of a local similarity measure. The knowledge engineer can use data distributions (left of Figure \ref{fig:age-sim}) to define the characteristics of the similarity measure.

In the following step the global similarity measures -- the weights for each particular attribute -- are defined. This can either be driven by knowledge, derived from data or learned over time. Especially the reduction or expansion of the variables available in the dataset is part of this feature engineering process.

\subsection{Retrieval}

\begin{figure}[h]
  \centering
  \includegraphics[width=0.47\textwidth]{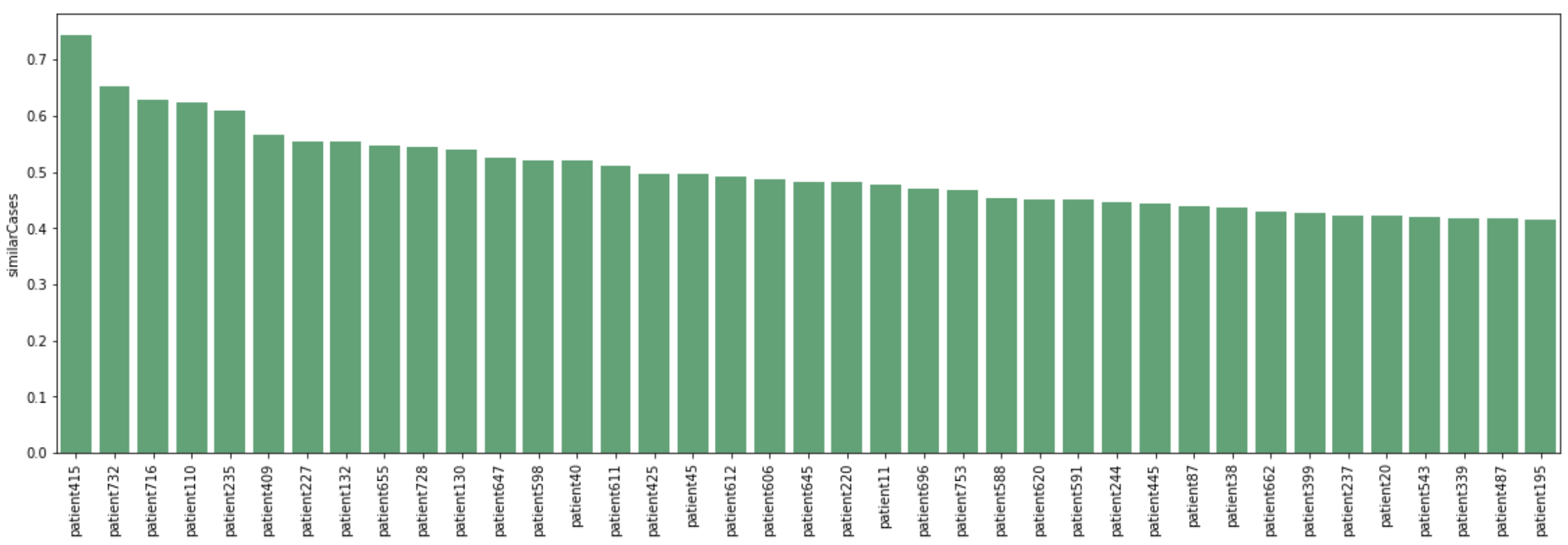}
  \caption{Retrieval results from the diabetes CBR system}
  \label{fig:retrievalresult}
\end{figure}

Once all similarity measures are defined, the case retrieval can be tested. The result of the CBR retrieval is usually a list of case-similarity pairs passed on to the adaptation engine and eventually presented to the user. For an example query to the diabetes dataset the result looks as shown in Figure \ref{fig:retrievalresult}.

 There are different approaches on how many cases are selected for further processing. During development of CBR systems the knowledge engineer(s) often need to inspect the details of the similarity functions comparing the cases to verify correct behaviour. 

\section{Similarity Visualizations}
\label{sec:vis}
In this section we present how retrieval insights can be presented to a knowledge engineer to gain understanding whether the similarity-based comparison is carried out as expected.

\begin{figure}[h]
  \centering
  \includegraphics[width=0.47\textwidth]{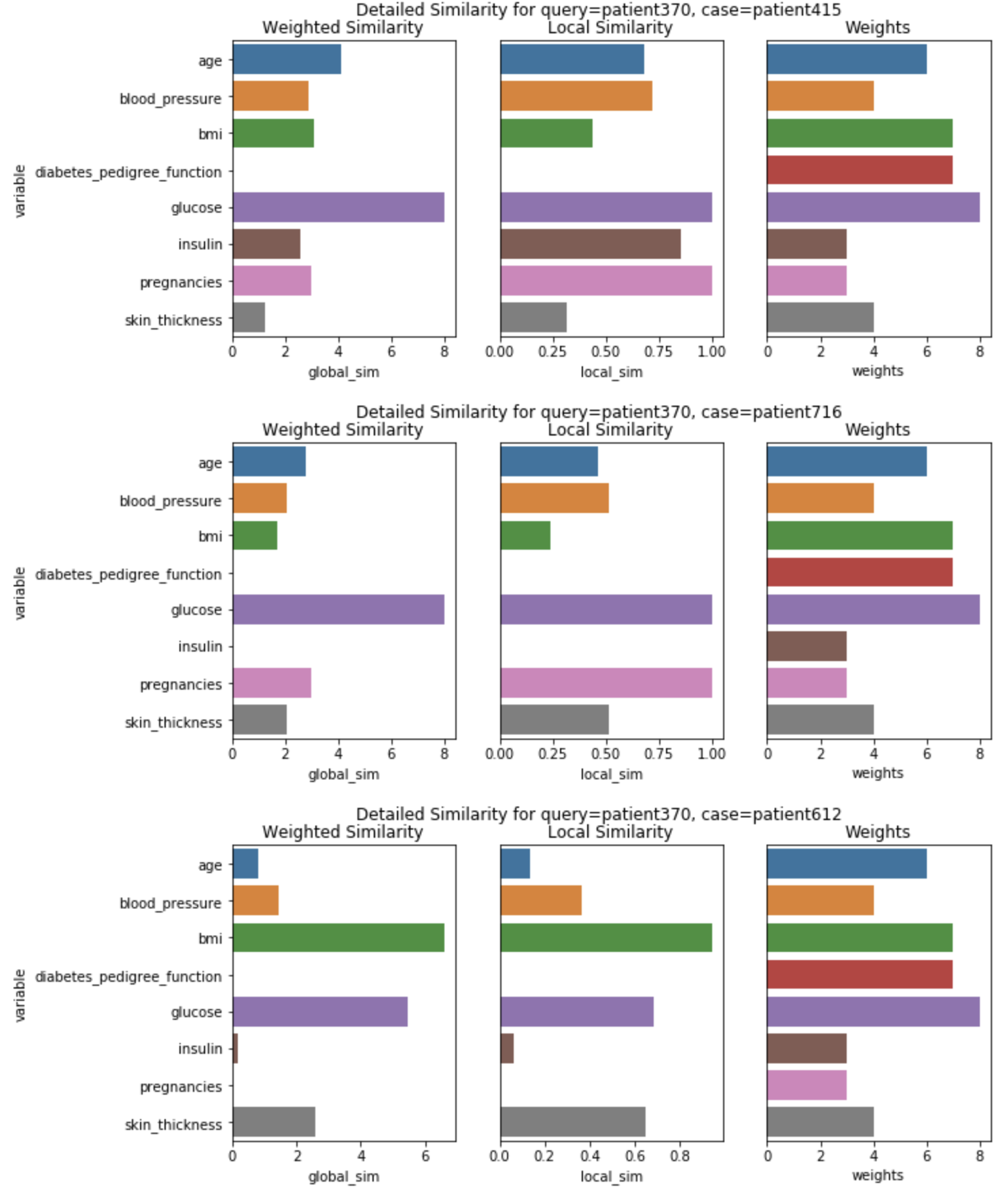}
  \caption{Visualization of weighted similarity, local similarity and weights}
  \label{fig:LocalSimViz}
\end{figure}

Figure \ref{fig:LocalSimViz} is an example of how the similarity scores leading up to the overall similarity presented in Figure \ref{fig:retrievalresult}. Each row of charts represents a comparison of a case from the case base to a query. The first row is the most similar case (highest ranked), followed by the second and third. The three charts are build the same way with the y-axis showing the attributes and the bars the similarity. The left row shows the weighted similarity score, the middle the similarity scores from each local similarity measure and the third chart shows the weights.

In the particular example given, one can see that the second and third case which have a similarity score of $0.62$ and $0.60$ respectively reach that score through different attributes. For the second case mostly the glucose levels are matching while for the third a lower glucose, but higher BMI. Such insights are certainly important for the development phase of the CBR system.

\section{Conclusion and Outlook}
\label{sec:summary}
In this paper we presented on similarity measures can be explained during the development process of CBR systems. As our work is mainly carried out in interdisciplinary teams, the current transparency was not explanatory enough. Therefore we found visualisations that help to understand the retrieval and can be tested using the CBR engine. All visualizations presented have been implemented using the \mcbr{} \rapi{} and python, matplotlib and seaborn for data handling and visualization. 

The next steps are to expand the visualizations towards the adaptation and case based evolution to gain better understanding when and how a CBR system is learning. With growing case bases, visualizing footprint cases and their provenance in the case base will also be in our focus.

\bibliography{bibliography}

\end{document}